\documentstyle[amsfonts,emlines,aps,preprint,epsfig,floats]{revtex}
\begin{document}
%%%%%%%%%%%%%

\title{ Parameter Symmetry of the Interacting Boson Model}
\author{
Andrey M.\ Shirokov,}
\address{
Institute for Nuclear Physics, Moscow State University,
Moscow 119899, Russia\thanks{Permanent address.}\\
and\\
International Institute for Theoretical and Applied Physics, \\
Iowa State University, Ames, Iowa 50011\\
E-mail: shirokov@anna19.npi.msu.su }
\author{N.\ A.\ Smirnova,}
\address{
Institute for Nuclear Physics, Moscow State University,
Moscow 119899, Russia\\
E-mail: smirnova@anna19.npi.msu.su}
%\author{and}
\author{%and\\
Yu.\ F.\ Smirnov,
}
\address{
Instituto de Ciencias Nucleares, Universidad Nacional Aut\'{o}noma de
M\'{e}xico,\\
 M\'{e}xico 04510, D.F., M\'{e}xico\\
and\\
Institute for Nuclear Physics, Moscow State University,
Moscow 119899, Russia\\
E-mail: smirnov@nuclecu.unam.mx}
%\date{\today }
%\date{ }
%%%%%%%%%%%%%%%%%%%%%%

\setcounter{footnote}{0}
\maketitle
\begin{abstract}
We discuss the symmetry of the parameter space of the interacting
boson model (IBM).
It is shown that for any set of the IBM Hamiltonian parameters
(with the only exception of the U(5) dynamical symmetry limit) one
can always find another set that generates the equivalent spectrum.
We discuss the origin of the symmetry and its relevance for physical
applications. \\
Pacs: 21.60.Fw \\
\end{abstract}
\pagebreak

In this paper we shall consider the simplest version of the interacting
boson model (\mbox{IBM-1}) that was proposed by Arima and Iachello~\cite
{Iachello1,Iachello2,Iachello3,Iachello4,Iachello} for the description of
spectra of low-lying states of even-even nuclei. IBM is now widely used in
nuclear applications. The symmetry properties of IBM that are determined by
the structure of the group U(6) have been thoroughly studied (see, e.g.,~ 
\cite{Iachello1,Iachello2,Iachello3,Iachello4,Iachello,Frank} and references
therein).

Below we shall discuss a symmetry in the space of the parameters of the IBM
Hamiltonian. We consider an irreducible set of parameters which because
of
the structure of the Lie algebra U(6), cannot be reduced to a smaller 
set~\cite{Iachello1,Iachello2,Iachello3,Iachello4,Iachello}. As we shall show,
the parameters in the IBM Hamiltonian can be transformed linearly
in such a way that the eigenvalue spectrum remains unchanged. 
We shall refer to this as parameter symmetry (PS). To the best of our
knowledge, the 
parameter symmetry has not been discussed before.

We shall show that PS can be related to ambiguities of the definition of
boson operators within IBM or, equivalently, to the possibility of using
different realizations of the SU(3) or SO(6) subalgebras of the U(6)
algebra. The ambiguities have been discussed
by various authors (see, e.~g., \cite{Iachello,Frank,VanIsacker,GDR}).
However, previous studies were mostly restricted to the cases of SU(3) and
SO(6) dynamical symmetry limits of IBM. Our approach is more general and we
emphasize manifestations of the transformations of the boson operators in
the space of parameters.
We make use of unitary transformations producing the
same (formal) effect on a Hamiltonian as  transformations of parameters.
This was also the basis of Refs.~\cite{Pauli,Pursey}
devoted to the weak interaction theory. The general
principle
%, {\em not restricted to the weak interaction application}, 
was carefully discussed in the review \cite{K-P-Pursey} where it was
called ``form invariance''.

The IBM Hamiltonian parameters are obtained by fitting the predictions of
the model to nuclear spectra known from the experiment.
Because of PS, the fit appears to be
essentially ambiguous. We discuss how to resolve the ambiguity of the fit in
applications.

Within IBM, nuclear states are labelled by a fixed number $N$ of bosons of
two types, $s$ and $d$, with quantum numbers $J^{\pi }=0^{+}$ and $J^{\pi
}=2^{+}$, respectively~\cite{Iachello1}. The six boson creation operators ($%
s^{+},d_{\mu }^{+},\;\mu =0,\pm 1,\pm 2$) and six boson annihilation
operators ($s,d_{\mu },\;\mu =0,\pm 1,\pm 2$) satisfy standard boson
commutation relations. The structure of the model is determined by the Lie
algebra U(6) generated by  36 bilinear combinations of these boson
operators~\cite{Iachello1,Iachello,Frank}.

The IBM Hamiltonian $H$ can be expressed as a superposition of the first ($%
C_{1}$) and second ($C_{2}$) order Casimir operators of the groups entering
the following reduction chains of the U(6) group~\cite
{Iachello1,Iachello,Frank} 
\begin{equation}
\label{d}
{\renewcommand{\arraystretch}{0}
\begin{array}{r@{}lcl@{\qquad}c}
& & \lefteqn{
\mbox{U(5)}\supset \mbox{SO(5)}\supset
\mbox{SO(3)}\supset \mbox{SO(2)}  }
&  & \qquad \mbox{I} \\
%& \nearrow &  \\
&\lefteqn{
\unitlength=.6mm
\special{em:linewidth 0.4pt}
\linethickness{0.4pt}
\begin{picture}(8.00,8.00)
\put(3.50,5.50){\oval(5.00,5.00)[rt]}
\emline{5.80}{5.60}{1}{3.80}{3.60}{2}
\emline{3.50}{8.00}{3}{1.50}{6.00}{4}
\end{picture} }\\
{\rm U}(6&) & %\rightarrow
\supset & \mbox{SU(3)}\supset \mbox{SO(3)}\supset
\mbox{SO(2)}& \qquad \mbox{II} \\
%& \searrow &  \\
&\lefteqn{
\unitlength=.6mm
\special{em:linewidth 0.4pt}
\linethickness{0.4pt}
\begin{picture}(8.00,8.00)
\put(3.50,0.50){\oval(5.00,5.00)[rb]}
\emline{5.85}{0.75}{5}{3.85}{2.75}{6}
\emline{1.40}{0.25}{7}{3.4}{-1.75}{8}
\end{picture} }\\
& & \lefteqn{
 \mbox{SO(6)}\supset \mbox{SO(5)}\supset \mbox{SO(3)}\supset
\mbox{SO(2)}   }
&  & \qquad  \mbox{III} \\
\end{array} }
\end{equation}
i. e. 
\begin{equation}
\begin{array}{r@{\ }l}
H(\{k_{i}\})= & H_{0}+k_{1}C_{1}(\mbox{U(5)})+k_{2}C_{2}(\mbox{U(5)})+\
k_{3}C_{2}(\mbox{SO(5)}) \\ 
& +\ k_{4}C_{2}(\mbox{SO(3)})+k_{5}C_{2}(\mbox{SO(6)})+k_{6}C_{2}(%
\mbox{SU(3)}),
\end{array}
\label{e}
\end{equation}
where $\{k_{i}\}\equiv \{H_{0},k_{1},k_{2},k_{3},k_{4},k_{5},k_{6}\}$ is the
set of the Hamiltonian parameters. The parameter set $\{k_{i}\}$ is
irreducible and  the number of parameters cannot be
reduced~\cite{Iachello1,Iachello}.  We define the Casimir operators
in Eq.~(\ref{e}) as in Ref.~\cite{Frank}. 

The main result of this letter can be stated as follows:

{\it Hamiltonians $H(\{k_{i}\})$ and $H(\{k_{i}^{\prime }\})$ defined by
Eq.~(\ref{e}) have identical eigenvalue spectra if the corresponding
irreducible parameter sets $\{k_{i}\}$ and $\{k_{i}^{\prime }\}$ are related
by 
\begin{equation}
\begin{array}{l}
H_{0}^{\prime }=H_{0},\ k_{1}^{\prime }=k_{1}+2k_{6},\ k_{2}^{\prime
}=k_{2}+2k_{6},\ k_{3}^{\prime }=k_{3}-6k_{6}, \\ 
k_{4}^{\prime }=k_{4}+2k_{6},\ k_{5}^{\prime }=k_{5}+4k_{6},\ k_{6}^{\prime
}=-k_{6}
\end{array}
\label{i1}
\end{equation}
in the case $k_{6}\neq 0$, or by 
\begin{equation}
\begin{array}{l}
H_{0}^{\prime }=H_{0}+10k_{5}N,\ k_{1}^{\prime }=k_{1}+4k_{5}(N{+}2),\
k_{2}^{\prime }=k_{2}-4k_{5}, \\ 
k_{3}^{\prime }=k_{3}+2k_{5},\ k_{4}^{\prime }=k_{4},\ k_{5}^{\prime
}=-k_{5},\ k_{6}^{\prime }=0
\end{array}
\label{i2}
\end{equation}
in the case $k_{6}=0$}.

To derive Eqs.~(\ref{i1})--(\ref{i2}) we consider the Hamiltonian matrix in
the basis $|Nnvn_{\Delta }LM\rangle $ (see, e.~g., \cite{Iachello})
associated with the reduction chain I in Eq.~(\ref{d}). The only quantum
number labelling the basis which is essential for the following discussion
is the number $n$ of $d$-bosons. Therefore we introduce the shortened
notation $\langle n|H|n^{\prime }\rangle \equiv 
\langle Nnvn_{\Delta}LM|H|Nn^{\prime}v^{\prime}n_{\Delta}^{\prime}LM\rangle$.  
The total number of bosons
labelling the totally symmetric representation of the U(6) group $N{=}n_{s}{+%
}n$ where $n_{s}$ is the number of $s$ bosons.

Obviously, the matrices of Casimir operators $C_{1}(\mbox{U(5)})$, $C_{2}(%
\mbox{U(5)})$, $C_{2}(\mbox{SO(5)})$ and $C_{2}(\mbox{SO(3)})$ of the groups
entering the reduction chain I contribute only to diagonal matrix elements $%
\langle n|H|n\rangle $. Off-diagonal matrix elements $\langle n|H|n^{\prime
}\rangle $ arise from the operators $C_{2}(\mbox{SO(6)})$ and $C_{2}(%
\mbox{SU(3)})$, defined by~\cite{Frank}, 
\begin{eqnarray}
&&C_{2}(\mbox{SO(6)})=N(N+4)-4(P^{+}{\cdot }\,P),  \label{CasimirSU} \\
&&C_{2}(\mbox{SU(3)})=2(Q\cdot Q)+\frac{3}{4}(L\cdot L),
\end{eqnarray}
where the multipole operators $P$, $L$, and $Q$ are
\begin{eqnarray}
&&P=\frac{1}{2}\left( (\tilde{d}\,\cdot \,\tilde{d})-(s\,\cdot \,s)\right) ,
\label{operatorsP} \\
&&L=\sqrt{10}\;[d^{+}{\times }\,\tilde{d}]^{(1)},
\label{operatorsL} \\
&&Q=[d^{+}{\times }\,s\ +\ s^{+}{\times }\,\tilde{d}]^{(2)}-\frac{\sqrt{7}}{2%
}[d^{+}{\times }\,\tilde{d}]^{(2)}, \label{operatorsQ}
\end{eqnarray}
with $\tilde{d}_{\mu }=(-1)^{\mu }d_{-\mu }$, ${\left( t^{(\lambda )}\cdot
u^{(\lambda )}\right) }=(-1)^{\lambda }\sqrt{2\lambda +1}\left[ t^{(\lambda
)}\times u^{(\lambda )}\right] ^{(0)}$, and ${\left[ t^{(\lambda
_{1})}\times u^{(\lambda _{2})}\right] _{\mu }^{(\lambda )}}%
=\sum\limits_{\mu _{1}\mu _{2}}(\lambda _{1}\mu _{1}\lambda _{2}\mu
_{2}|\lambda \mu )\,t_{\mu _{1}}^{(\lambda _{1})}u_{\mu _{2}}^{(\lambda
_{2})}$ where $(\lambda _{1}\mu _{1}\lambda _{2}\mu _{2}|\lambda \mu )$ is a
Clebsch-Gordan coefficient. The operator $C_{2}(\mbox{SU(3)})$ contributes
to the off-diagonal matrix elements $\langle n|H|n\pm 1\rangle $ and $%
\langle n|H|n\pm 2\rangle $ while the operator $C_{2}(\mbox{SO(6)})$
contributes to off-diagonal matrix elements only of the type $\langle
n|H|n\pm 2\rangle $. Both Casimir operators also contribute to the diagonal
matrix elements $\langle n|H|n\rangle $. Thus in the general case the matrix
of the Hamiltonian (\ref{e}) is five-diagonal.

Let us now transform the matrix $\langle n|H|n^{\prime }\rangle $ using the
unitary transformation $\langle n|U_{1}|n^{\prime }\rangle =(-1)^{n}\delta
_{nn^{\prime }}$ that produces a transformed $H^{\prime }=(U_{1})^{-1}HU_{1}$
differing from $H$ only by the sign of the off-diagonal matrix elements $%
\langle n|H|n\pm 1\rangle $. Since the only Casimir operator which
contributes to these matrix elements is $C_{2}(\mbox{SU(3)})$, the sign of $%
\langle n|H|n\pm 1\rangle $ can be restored by setting $k_{6}\rightarrow
k_{6}^{\prime }=-k_{6}$. The diagonal matrix elements $\langle n|H|n\rangle $
and the off-diagonal ones of type $\langle n|H|n\pm 2\rangle $ may then be
restored by setting $k_{i}\rightarrow k_{i}^{\prime }$ for all the rest of
the parameters, where the $k_{i}^{\prime }$ were defined in (\ref{i1}). This
claim is easily verified using the explicit expressions for the Casimir
operators found elsewhere \cite{Frank}.

When $k_{6}=0$, the only non-zero matrix elements are $\langle n|H|n\rangle $
and ${\langle n|H|n\pm 2\rangle }$, so that the Hamiltonian matrix is
tridiagonal. Thus we can use the unitary transformation $U_{2}$ that
transforms the Hamiltonian into $H^{\prime }=(U_{2})^{-1}HU_{2}$ with
non-zero matrix elements $\langle n|H^{\prime }|n\rangle =\langle
n|H|n\rangle $ and $\langle n|H^{\prime }|n\pm 2\rangle =-\langle n|H|n\pm
2\rangle $. The sign change of the off-diagonal matrix elements $\langle
n|H|n\pm 2\rangle $ can be restored by setting $k_{5}\rightarrow
k_{5}^{\prime }=-k_{5}$. To restore the diagonal matrix elements, we
redefine the parameters $k_{i}$, $1\leq i\leq 4$ according to (\ref{i2}).%

Thus we have shown that for any irreducible set of the IBM Hamiltonian
parameters there is another irreducible set which yields the same spectrum.
The only exception is the U(5) dynamical symmetry (DS) limit 
($k_{5}=k_{6}=0$) when the two sets of parameters coincide, as is easily seen
from (\ref{i2}).

An intriguing consequence of PS is that it establishes an equivalence
between the nuclear spectrum corresponding to a certain DS and a
transitional nuclear spectrum. As follows from Eqs.~(\ref{i1}), the
rotational spectrum of the SU(3) DS limit ($k_{1}{=}k_{2}{=}k_{3}{=}k_{5}{=}0
$) is identical to the spectrum of the transitional Hamiltonian with the
parameters 
\begin{equation}
\{k_{i}^{\prime }\}\equiv \left\{ k_{1}^{\prime }{=}2k_{6},\ k_{2}^{\prime }{%
=}2k_{6},\ k_{3}^{\prime }{=}{-}6k_{6},\ k_{4}^{\prime }{=}k_{4}{+}2k_{6},\
k_{5}^{\prime }{=}4k_{6},\ k_{6}^{\prime }{=}{-}k_{6}\right\} ,
\label{primeSU3}
\end{equation}
which does not correspond to any DS. Similarly, it follows from (\ref{i2})
that the $\gamma $-unstable spectrum of the SO(6) DS limit ($k_{1}{=}k_{2}{=}%
k_{6}{=}0$) is identical to the \mbox{U(5)--SO(6)} transitional nuclear
spectrum with the parameters
\begin{equation}
\{k_{i}^{\prime }\}\equiv \left\{ k_{1}^{\prime }{=}4(N{+}2)k_{5},\
k_{2}^{\prime }{=}{-}4k_{5},\ k_{3}^{\prime }{=}k_{3}{+}2k_{5},\
k_{4}^{\prime }{=}k_{4},\ k_{5}^{\prime }{=}{-}k_{5},\ k_{6}^{\prime }{=}%
0\right\} .  \label{primeSO6}
\end{equation}

To understand the origin of the PS discussed above, we note that there is an
ambiguity in definition of boson operators within IBM~\cite
{Iachello,Frank,VanIsacker}. For example, one can change the sign of the
creation $s^{+}$ and annihilation $s$ operators without violating the boson
commutation relations. Obviously, the transformation 
\begin{equation}
\left\{ s^{+}\rightarrow -s^{+},\ \ s\rightarrow -s\right\}  \label{sminuss}
\end{equation}
should not produce any change of the spectra. From 
Eqs.~(\ref{CasimirSU})--(\ref{operatorsQ}) it is seen that only the
Casimir operator $C_{2}(\mbox{SU(3)})$ is changed under the
transformation (\ref{sminuss}) [we note 
that the Casimir operators $C_{1}(\mbox{U(5)})$, $C_{2}(\mbox{U(5)})$, 
$C_{2}(\mbox{SO(5)})$ and $C_{2}(\mbox{SO(3)})$ can be expressed through
bilinear combinations of $d$ boson operators only and are unchanged under
any transformation of $s$ boson operators]. The transformation 
(\ref{sminuss}) is equivalent to the transformation of the quadrupole
operator $Q\rightarrow \overline{Q}$ where 
\begin{equation}
\overline{Q}=-[d^{+}{\times }\,s\,+\,s^{+}{\times }\,\tilde{d}]^{(2)}-\frac{%
\sqrt{7}}{2}[d^{+}{\times }\,\tilde{d}]^{(2)}.  \label{Q}
\end{equation}
It is clear from Eqs.~(\ref{CasimirSU})--(\ref{operatorsQ}) and (\ref{Q})
that the off-diagonal matrix elements $\langle n|H|n\pm 1\rangle $ change
their sign under the transformation (\ref{sminuss}) while all the remaining
matrix elements of the Hamiltonian remain unchanged. Thus, the PS
transformation (\ref{i1}) restores the original form of the IBM
Hamiltonian 
$H$ subjected to the transformation (\ref{sminuss}), or, in other
words, the PS
transformation (\ref{i1}) is equivalent to the transformation (\ref{sminuss}).

The quadrupole operators $Q$ and $\overline{Q}$ correspond to different
embeddings of the SU(3) subgroup in the U(6) group (see also \cite{Dieperink}
for other realizations of SU(3)). The Casimir operator $C_{2}\!\left( 
\overline{\mbox{SU(3)}}\right) $ of the $\overline{\mbox{SU(3)}}$ algebra
associated with the quadrupole operator (\ref{Q}) can be expressed through $%
C_{2}(\mbox{SU(3)})$ and other Casimir operators using (\ref{primeSU3}) with 
$k_{4}=0$ and $k_{6}=1$: 
%\begin{equation}
%\label{CasimirSU3}
%C_2(\mbox{SU$^{\prime }$(3)})=2(\hat Q^{\prime }\cdot \hat Q^{\prime })
%+\frac34(\hat L\cdot \hat L),
%\end{equation}
%has the same matrix elements as $C_2(\mbox{SU(3)})$, except of those
%contributing to $\langle n|H^{\prime }|n\pm 1\rangle $.
%$C_2(\mbox{SU$^{\prime }$(3)})$ can be expressed through
%$C_2(\mbox{SU(3)})$ and other Casimir operators as
\begin{equation}
\begin{array}{rl}
C_{2}\!\left( \overline{\mbox{SU(3)}}\right) = & 2C_{1}(\mbox{U(5)})+2C_{2}(%
\mbox{U(5)})-6C_{2}(\mbox{SO(5)}) \\ 
& +\ 2C_{2}(\mbox{SO(3)})+4C_{2}(\mbox{SO(6)})-C_{2}(\mbox{SU(3)}),
\end{array}
\label{m1}
\end{equation}

Similarly, in the case $k_{6}=0$ one can change the sign of bilinear
combinations $\left( s^{+}{\cdot }\,s^{+}\right) $ and $\left( s\cdot
s\right) $ and, consequently, the sign of off-diagonal matrix elements $%
\langle n|H|n\pm 2\rangle $, by using the transformation 
\begin{equation}
\left\{ s^{+}\rightarrow \mbox{i}s^{+},\ \ s\rightarrow -\mbox{i}s\right\} .
\label{sis}
\end{equation}
The IBM Hamiltonian $H$ subjected to the transformation (\ref{sis}) can be
restored to its original form using the PS transformation (\ref{i2}). On the
other hand, in the case $k_{6}=0$ when $C_{2}(\mbox{SU(3)})$ is not present
in the Hamiltonian, the transformation (\ref{sis}) is equivalent to the
transformation of the monopole operator $P\rightarrow \overline{P}$ where 
\begin{equation}
\overline{P}=\frac{1}{2}\left( (\tilde{d}\cdot \tilde{d})+(s\cdot s)\right) .
\label{P}
\end{equation}
This monopole operator is associated with an alternative embedding of the
SO(6) subgroup in the U(6) group \cite{Frank,VanIsacker,GDR}. By using (\ref
{primeSO6}) with $k_{3}=k_{4}=0$ and $k_{5}=1$, we find the Casimir operator
of the $\overline{\mbox{SO(6)}}$ algebra associated with the monopole
operator $\overline{P}$ to be 
\begin{equation}
C_{2}\!\left( \overline{\mbox{SO(6)}}\right) =10N+4(N{+}2)\,C_{1}(\mbox{U(5)}%
)-4C_{2}(\mbox{U(5)})+2C_{2}(\mbox{SO(5)})-C_{2}(\mbox{SO(6)})\,.  \label{n1}
\end{equation}

Are the parameter symmetries (\ref{i1}) and (\ref{i2}) the only ones
present in IBM? We have shown that 
these parameter symmetries are associated with the phase ambiguity of the
boson operators. The general phase transformation of boson operators
consistent with standard boson commutation relations is \cite
{VanIsacker,Iachello} 
\begin{equation}
\left\{ b^{+}\rightarrow e^{\mbox{\scriptsize i}\varphi }b^{+},\ \
b\rightarrow e^{-\mbox{\scriptsize i}\varphi }b\right\} .  \label{seifis}
\end{equation}
However time reversal symmetry implies severe restrictions on the values of $%
\varphi $ \cite{VanIsacker,Frank}, namely $\varphi =0,\;\pi $ in the case of
the general IBM Hamiltonian, and $\displaystyle\varphi =0,\;\pm \frac{\pi }{2%
},\;\pi $ in the case of the transitional SO(6)--U(5) IBM Hamiltonian with $%
k_{6}=0$. It is easy to check that we do not obtain new parameter symmetries
using all possible phase transformations of $s$, or $d$, or both $s$ and $d$
boson operators. Parameter symmetries also cannot be
generated by spatial rotations. Thus by using all transformations of boson
operators consistent with $N$ and angular momentum conservation and the time
reversal symmetry one can obtain only the parameter symmetries (\ref{i1})
and (\ref{i2}). Even the particle-hole boson
transformation discussed in Refs.~\cite{Dieperink,Frank}, which is not $N$%
-conserving but is nevertheless isospectral for some particular IBM
Hamiltonians, does not generate new parameter symmetries.

However our approach based on the study of the structure of the Hamiltonian
matrix and its unitary transformations is more general than that based on
the transformation of boson operators. Therefore it is possible to derive an
additional PS that cannot be formulated in terms of boson transformations.
The Hamiltonian $\overline{H\left( \{\overline{k}_{i}\}\right) }$ defined by
replacing $C_{2}(\mbox{SO(6)})$ in (\ref{e}) by $C_{2}\!\left( \overline{%
\mbox{SO(6)}}\right) $ is identical to and therefore isospectral with
$H(\{k_{i}\})$ if 
\begin{equation}
\begin{array}{l}
\overline{H}_{0}=H_{0}+10k_{5}N,\ \overline{k}_{1}=k_{1}+4k_{5}(N{+}2),\ 
\overline{k}_{2}=k_{2}-4k_{5}, \\ 
\overline{k}_{3}=k_{3}+2k_{5},\ \overline{k}_{4}=k_{4},\ \overline{k}%
_{5}=-k_{5},\ \overline{k}_{6}=k_{6}.
\end{array}
\label{ibar}
\end{equation}
One can combine Eqs.~(\ref{ibar}) and (\ref{i1}) to get one more PS
relation. 
The quasiclassical approach of Ref.~\cite{Fursa} applied to the %IBM
Hamiltonian (\ref{e}) does not permit additional parameter symmetries 
%of IBM. However 
but we do not have a formal proof of the
absence of additional parameter symmetries of IBM. However it is very
probable that  additional parameter symmetries not
associated with transformations of boson operators, can be found in
some more complicated models than IBM-1, for example in IBM-2.

Let us discuss whether there is a possibility of discriminating between the
two sets of IBM parameters which gives rise to equivalent spectra. For this
purpose it is natural to consider electromagnetic transitions which were
studied within IBM in a number of papers (see~\cite{E2} and references
therein).  In the Consistent-$Q$ Formalism (C$Q$F) of Warner and
Casten~\cite{Warner},  the quadrupole operator 
\begin{equation}
Q^{\chi }=[d^{+}{\times }\,s\ +\ s^{+}{\times }\,\tilde{d}]^{(2)}+\chi
\lbrack d^{+}{\times }\,\tilde{d}]^{(2)}  \label{QHI}
\end{equation}
is introduced
instead of operators $P$ and $Q$, and both $(P^{+}{\cdot }\,P)$
and $(Q\cdot Q)$ terms in the IBM Hamiltonian are replaced by a single
term $(Q^{\chi }{\cdot }\,Q^{\chi })$. The operator (\ref{QHI}) should
be used for calculations of $E2$ transition rates. The transformation
(\ref{sminuss}) changes $Q^{\chi }$ into the operator 
\begin{equation}
\overline{Q^{\chi }}=-[d^{+}{\times }\,s\ +\ s^{+}{\times }\,\tilde{d}%
]^{(2)}+\chi \lbrack d^{+}{\times }\,\tilde{d}]^{(2)}=-Q^{-\chi }.
\label{QHIbar}
\end{equation}
The consistent transformation of the $E2$ transition operator $Q^{\chi
}\rightarrow \overline{Q^{\chi }}$ and of the quadrupole-quadrupole
interaction $(Q^{\chi }{\cdot }\,Q^{\chi })\rightarrow \left( \overline{%
Q^{\chi }}{\cdot }\,\overline{Q^{\chi }}\right) $ in the Hamiltonian
guarantees that the $E2$ transition rates remain unchanged. Note  
that in the general
case, i.~e. when $k_{6}\neq 0$, the parameter $\chi $ is nonzero and
the quadrupole-quadrupole interaction $(Q^{\chi }{\cdot }\,Q^{\chi })$ is
unambiguously determined by the set of the Hamiltonian parameters
$\{k_{i}\}$. Thus PS (\ref{i1}) 
establishes the isospectrality of the Hamiltonian $H$ with the one in
which the quadrupole-quadrupole interaction $(Q^{\chi }{\cdot }\,Q^{\chi })$
term is replaced by the term $\left( \overline{Q^{\chi }}{\cdot }\,\overline{%
Q^{\chi }}\right) $. $E2$ transitions cannot be used to distinguish between
the two Hamiltonians within the formalism of C$Q$F.

However the statement that one cannot use the operator $\overline{Q^{\chi
}}$ for calculation of $E2$ transitions in the case when the
quadrupole-quadrupole interaction entering the Hamiltonian is of the form $%
(Q^{\chi }{\cdot }\,Q^{\chi })$, is somewhat doubtful. The operators $%
(Q^{\chi }{\cdot }\,Q^{\chi })$ and $\left( \overline{Q^{\chi }}{\cdot }\,%
\overline{Q^{\chi }}\right) $ (and the corresponding Hamiltonians) are
isospectral and their matrices differ by the sign of off-diagonal elements
only. Therefore it would be interesting to study $E2$ transitions using both
operators (\ref{QHI}) and (\ref{QHIbar}), and to learn whether the
phenomenological transition rates follow the prescriptions of C$Q$F.

The case $k_{6}=0$ is much more interesting and complicated. In this case
one should substitute $\chi $ by 0 in (\ref{QHI}). We note that $Q^{0}$ is a
generator of SO(6). Under the transformation (\ref{sis}), $Q^{0}$ becomes 
\begin{equation}
\overline{Q^{0}}=-\mbox{i}[d^{+}{\times }\,s\ -\ s^{+}{\times }\,\tilde{d}%
]^{(2)}  \label{Q0bar}
\end{equation}
(we note that (\ref{Q0bar}) presents an alternative form of the SO(6)
generator \cite{Frank}). In contrast to the case $\chi \neq 0$, the
interpretation of the quadrupole-quadrupole interaction in the Hamiltonian
in terms of the $(Q^{0}{\cdot }\,Q^{0})$ or $\left( \overline{Q^{0}}{\cdot }%
\,\overline{Q^{0}}\right) $ operator is now ambiguous. It is easy to show
that 
\begin{equation}
(Q^{0}{\cdot }\,Q^{0})=-\left( \overline{Q^{0}}{\cdot }\,\overline{Q^{0}}%
\right) +10N+4(N{+}2)\,C_{1}(\mbox{U(5)})-4C_{2}(\mbox{U(5)})-2C_{2}(%
\mbox{SO(5)})\,.  \label{Q0ambig}
\end{equation}
Thus in the case $k_{6}=0$ the IBM Hamiltonian can be expressed either
through $(Q^{0}{\cdot }\,Q^{0})$ or alternatively through $\left( \overline{%
Q^{0}}{\cdot }\,\overline{Q^{0}}\right) $. As a result, the definition of
the $E2$ transition operator appears to be ambiguous.
The operators $Q^{0}$ and $\overline{Q^{0}}$ provide different $E2$ rates
for some transitions \cite{VanIsacker} and identical rates for the remaining
ones. Thus it would be interesting to compare carefully the results of
calculations with $Q^{0}$ and $\overline{Q^{0}}$ with phenomenological data
on electromagnetic transitions in a number of SO(6)--U(5) transitional
nuclei.

We can interpret PS in the case $k_{6}=0$ as a phase transformation of the
boson operators (\ref{sis}). Then we should use different quadrupole
operators $Q^{0}$ and $\overline{Q^{0}}$ for the different Hamiltonians
connected by the PS. As a result, the two Hamiltonians cannot be
distinguished by the electromagnetic transitions. Alternatively, we can
interpret the PS (\ref{i2}) as a possibility of constructing two isospectral
Hamiltonians expressed through the same set of operators. Then we should use
the same quadrupole operator for the calculations of $E2$ transitions for
both Hamiltonians. As a result, we find that the two Hamiltonians can be
distinguished by the electromagnetic transitions. So, due to the ambiguity
in the definition of the quadrupole operator, there is no definite answer on
the question about the distinguishability of the PS-related Hamiltonian by
means of $E2$ transitions in the case $k_{6}=0$.

One might suggest a resolution of the ambiguity by requiring the
quadrupole-quadrupole interaction to be attractive and expressing the
interaction either by the operator $(Q^{0}{\cdot }\,Q^{0})$ or by the
operator $\left( \overline{Q^{0}}{\cdot }\,\overline{Q^{0}}\right) $
according to the sign of the term in the Hamiltonian. However this
prescription seems dubious. While changing the sign of the monopole-monopole
(pairing) interaction results in non-trivial physical issues (see below),
the change of the sign of the pure quadrupole-quadrupole interaction is not
manifested in observables. The two Hamiltonians related by the PS (\ref
{i1}) differ only by the sign of the quadrupole-quadrupole term $(Q\cdot Q)$
and are indistinguishable in physical applications at least within the
framework of C$Q$F. However the sign of the $(Q^{\chi }{\cdot }\,Q^{\chi })$
term in the Hamiltonian in the general case $\chi \neq 0$ is not arbitrary,
and is manifested in physical observables (see below). This is because the C$%
Q$F quadrupole-quadrupole interaction $(Q^{\chi }{\cdot }\,Q^{\chi })$
accounts effectively for the pairing interaction, and the sign of the term $%
(Q^{\chi }{\cdot }\,Q^{\chi })$ is determined mostly by the sign of the
pairing term $(P^{+}{\cdot }\,P)$ and not by the sign of the
quadrupole-quadrupole term $(Q\cdot Q)$.

There is another possibility for distinguishing between two PS-related
Hamiltonians in the case $k_{6}=0$. An intriguing feature of the PS (\ref{i2}%
) is its $N$-dependence. Therefore if it is supposed that
the spectra of neighboring even-even nuclei are described by the same set of
IBM parameters (see for example Ref.~\cite{Ndependence}), then one can
discriminate between the parameters connected by (\ref{i2}) comparing the
spectra of the neighboring nuclei. Since the relations (\ref{i2}) involve
the total number of bosons $N$, the two sets of the parameters can yield
identical spectra only for a particular nucleus, and will yield 
different ones for its isotopes or
isotones. In Fig.~1 we present the spectra of few isotopes of Pt. The set of
parameters $k_{1}=k_{2}=k_{6}=0$, $k_{3}=50$~keV, $k_{4}=10$~keV, $%
k_{5}=-42.75$~keV was suggested in \cite{Iachello4}
for the description of $^{196}$Pt ($N$=6) within the SO(6) DS limit of IBM.
The corresponding spectra are given in the left columns labelled by SO(6).
The set of parameters $k_{1}^{\prime }=-1368$~keV, $k_{2}^{\prime }=171$%
~keV, $k_{3}^{\prime }=-35.5$~keV, $k_{4}^{\prime }=10$~keV, $k_{5}^{\prime
}=42.75$~keV and $k_{6}^{\prime }=0$ is obtained using (\ref{i2}) with $N=6$%
. The corresponding spectra are given in the right columns labelled by PS.
The SO(6) and PS spectra are, of course, identical in the case of $^{196}$Pt
but differ for other Pt isotopes. The difference is seen to be
essential.  The SO(6) DS set of
parameters suggested in \cite{Iachello4} provides a reasonable description
of $^{192}$Pt and $^{194}$Pt although these isotopes were not involved in
the fit. The alternative set of parameters obtained using PS fails to
reproduce the spectra of $^{192}$Pt and $^{194}$Pt.
%
%
%
%%%%%%%%%%%%%%%%%%%%%%%
\begin{figure}
\epsfverbosetrue
\epsfxsize=1\textwidth
\epsfbox{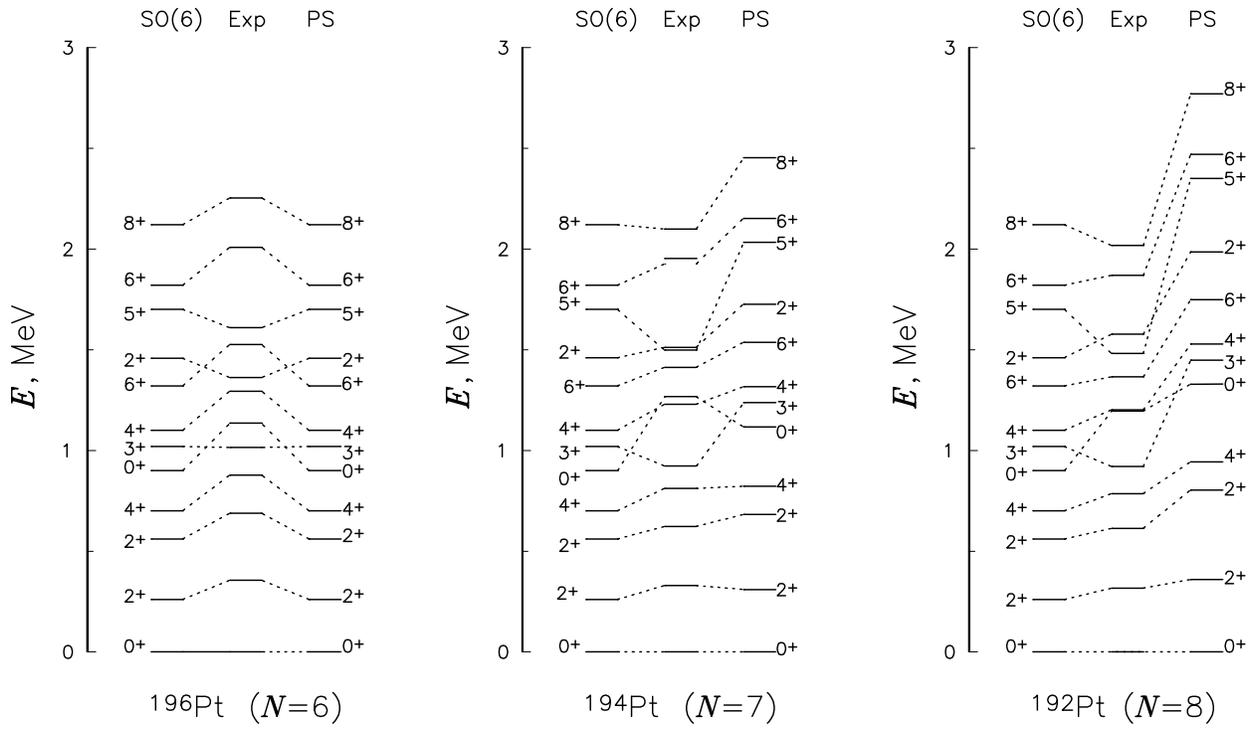}

\vspace{.6cm}

\caption{Few lowest levels of each $J^{\pi}$ of Pt isotopes. SO(6):
calculations within SO(6) DS limit with the parameters
suggested in \protect\cite{Iachello4} for the description of $^{196}$Pt
($N=6$); PS: calculations with the set of parameters obtained using
(\protect\ref{i2}) with $N=6$; Exp: experimental data of
Ref.~\protect\cite{ExpPt}.}

\end{figure}
%%%%%%%%%%%%%%%%%%%%%%%%%%%%
%%
%

The spectra of Casimir operators 
$C_{2}\!\left(\overline{\mbox{SO(6)}}\right)$ and $C_{2}(\mbox{SO(6)})$ 
are, of course, identical for all values of $N$. However by expressing 
$C_{2}\!\left( \overline{\mbox{SO(6)}}\right) $ in terms of $C_{2}(%
\mbox{SO(6)})$ using (\ref{n1}), we obtain the difference in the  
$N$-dependence of the spectra.  This $N$-dependence results from the 
change in the sign of the pairing interaction: $C_{2}\!\left( 
\overline{\mbox{SO(6)}}\right) $ may be expressed in terms of 
the monopole operator $\overline{P}$ by replacing $(P^{+}{\cdot }\,P)$ 
with $\left( {\overline{P}%
\vphantom{\overline{\overline{P}}}}^{\,+}{\cdot }\,\overline{P}\right) $ in
Eq.~(\ref{CasimirSU}). Hence Eq.~(\ref{n1}) is just the expression of
the attractive (repulsive) pairing interaction $(P^{+}{\cdot }\,P)$ in
terms of the alternative repulsive (attractive) pairing interaction
$\left( {\overline{P}
\vphantom{\overline{\overline{P}}}}^{\,+}{\cdot }\,\overline{P}\right)$.
Equivalently, the PS (\ref{i2}) states the $N$-dependent isospectrality of
two IBM Hamiltonians with opposite signs of the pairing interaction. Thus,
in contrast to that of the quadrupole-quadrupole interaction, the sign
of the pairing 
interaction is not arbitrary and is manifested in the $N$-dependence of the
spectrum. As a result, we can exclude one of the PS-related
Hamiltonians by the requirement that the pairing interaction in a
physically-acceptable Hamiltonian should be attractive. This is illustrated
in Fig. 1.
The SO(6) DS Hamiltonian with the attractive pairing interaction fitted to
the $^{196}$Pt spectra in Ref.~\cite{Iachello4} also reproduces the spectra
of $^{192}$Pt and $^{194}$Pt, while the unphysical Hamiltonian with the
repulsive pairing term fails to do so.

The change of sign of the pairing interaction is also manifested in the $N$%
-dependence of the PS (\ref{ibar}). Thus the PS (\ref{ibar}) also
demonstrates the $N$-dependent isospectrality between a
physically acceptable IBM Hamiltonian and an unphysical one
with a repulsive pairing interaction.

In this paper we have shown that there is a symmetry in the parameter space
of IBM which manifests itself by two IBM Hamiltonians defined by different
irreducible parameter sets having identical spectra. Some of the
PS relations but not others can be associated with the ambiguities in the
definition of boson operators. We have shown that PS (\ref{i1}) relates two
physically indistinguishable IBM Hamiltonians. In our opinion, this 
indistinguishability originates in the arbitrariness of the sign of
the quadrupole-quadrupole interaction $(Q\cdot Q)$. In contrast, the
other parameter symmetries relate IBM Hamiltonians 
which have pairing interactions of opposite sign. The sign of the pairing
interaction is not arbitrary and its change manifests itself in physical
observables like the $N$-dependency of the spectrum. 
Thus these PS-related Hamiltonians can be distinguished
by studying the spectra of neighboring nuclei, and one of the two can be
excluded as unphysical. We have shown also that there is an ambiguity in the
definition of the quadrupole operator in the case of transitional
SO(6)--U(5) nuclei. We believe that
these results are of importance for applications, since the IBM
parameters are conventionally obtained by fitting experimentally known
nuclear spectra.

The phase transformations of boson operators of the type
(\ref{sminuss}) have been discussed in Ref.~\cite{Yoshida}.
An additional $d$-parity quantum number connected with 
this transformation was introduced in Ref.~\cite{d-parity} 
and used for classification of electromagnetic transitions 
in $\gamma $-unstable nuclei within IBM-2. 

IBM is not the only boson model to display parameter symmetries. In
particular, we have shown in Ref.~\cite{VM} that PS is present in the vibron
model~\cite{IachelloVM} describing the low-lying excitations of diatomic
molecules. Various parameter symmetries are surely present in extensions of
IBM-1, such as \mbox{IBM-2} which distinguishes proton and neutron degrees
of freedom, $sdg$-IBM, etc. However we have not yet performed a careful
study of all possible parameter symmetries of either IBM-2 or $sdg$-IBM. 
It would be also interesting to study parameter symmetries of fermion
models used in nuclear physics.

One non-trivial consequence of parameter
symmetries is that the energy spectrum of a DS limit of a boson model
can be exactly reproduced by an
irreducible set of parameters which does not correspond to any DS limit.
From a physical point of view this means, for example, 
that a typical rotational
nuclear spectrum can be reproduced by a combination of rotations and
vibrations. As a result, boson Hamiltonians demonstrate the 
so-called `hidden symmetries' that seem to be
important \cite{Kusnezov} in the studies of chaos and quantum
nonintegrability.

Parameter symmetries
originate from the possibility of using different realizations of SO(6) and
SU(3) algebras within IBM which correspond to different embeddings of SO(6)
and SU(3) subgroups in the U(6) group. Thus we used the parameter symmetries
(\ref{i1}) and (\ref{i2}) to derive expressions (\ref{m1}) and (\ref{n1})
of the Casimir operators 
$C_{2}\!\left(\overline{\mbox{SU(3)}}\right)$ and
$C_{2}\!\left(\overline{\mbox{SO(6)}}\right)$ of 
alternative algebras $\overline{\mbox{SU(3)}}$
and $\overline{\mbox{SO(6)}}$ in terms of Casimir operators 
$C_{2}(\mbox{SU(3)})$ and  $C_{2}(\mbox{SO(6)})$ of the original $\mbox{SU(3)}$
and $\mbox{SO(6)}$ algebras. These expressions are valid within the totally
symmetric irreducible representation of U(6).  
This approach seems to be very promising for generating
independent algebras in the case of more complicated boson models. For example,
we used it to construct the most general IBM-2 Hamiltonian~\cite{DS}.

%The idea of studying symmetry of the parameter space by means of the unitary
%transformation, has been exploited before~\cite{Pauli,Pursey} in the theory
%of four-fermion weak interaction. In particular, it was shown in Refs.~\cite
%{Pauli,Pursey} that all physical observables are governed by certain
%bilinear combinations of coupling constants which are invariant under
%unitary transformations.

We are thankful to 
R.~Bijker, D.~Bonatsos, O.~Casta\~nos, J.~Cseh, C.~Daskaloyannis,
G.~F.~Filippov, A.~Frank, F.~Iachello, R.~V.~Jolos,  
T.~Otsuka, J.~Patera, N.~Pietralla, V.~N.~Tolstoy and P.~van~Isacker
for encouraging discussions. We are also grateful to D.~L.~Pursey for
careful reading of the manuscript and valuable comments. One of the authors
(A.M.S.) is grateful to the International Institute for Theoretical and
Applied Physics at Iowa State University (ISU) for the hospitality and for
the financial support of his stay in ISU. The work was supported in part by
the Russian Foundation of Fundamental Research under the Grant No.
96-01-01421 and by the Competitive Center at St.~Petersburg State University.

%%%%%%%%%%%%%%%%%%%%%%%
%\begin{figure}
%\epsfverbosetrue
%\epsfxsize=1\textwidth
%\epsfbox{fig.eps}

%\vspace*{2cm}

%\centerline{FIGURE CAPTION}

%\vspace*{2cm}

%\caption{Few lowest levels of each $J^{\pi}$ of Pt isotopes. SO(6):
%calculations within SO(6) DS limit with the parameters
%suggested in \protect\cite{Iachello4} for the description of $^{196}$Pt
%($N=6$); PS: calculations with the set of parameters obtained using
%(\protect\ref{i2}) with $N=6$; Exp: experimental data of
%Ref.~\protect\cite{ExpPt}.}

%\end{figure}
%\newpage
%\addtocounter{figure}{-1}

%\begin{figure}
%\epsfverbosetrue
%\epsfxsize=1\textwidth
%\epsfbox{fig.eps}

%\vspace{2cm}

%\caption{A.~M.~Shirokov, N.~A.~Smirnova, and Yu.~F.~Smirnov,
%Phys. Lett. B.}

%\end{figure}
%%%%%%%%%%%%%%%%%%%%%%%%%%%%

\end{document}